\documentclass[letterpaper]{article} 
\usepackage{aaai23}  
\usepackage{times}  
\usepackage{helvet}  
\usepackage{courier}  
\usepackage[hyphens]{url}  
\usepackage{graphicx} 
\usepackage{natbib}  
\usepackage{caption} 
\usepackage{amsmath}
\usepackage{amssymb}
\usepackage{booktabs}
\usepackage{multirow}
\usepackage{makecell}
\usepackage{enumitem}
\usepackage[linesnumbered, ruled, longend]{algorithm2e}
\usepackage[title]{appendix}
\let\sss= \scriptscriptstyle

\title{Untargeted Attack against Federated Recommendation Systems\\via Poisonous Item Embeddings and the Defense}
\author {
    Yang Yu\textsuperscript{\rm 1,2}, Qi Liu\textsuperscript{\rm 1,2}\thanks{Corresponding Author.}, Likang Wu\textsuperscript{\rm 1,2}, Runlong Yu\textsuperscript{\rm 1,2}, Sanshi Lei Yu\textsuperscript{\rm 1,2}, Zaixi Zhang\textsuperscript{\rm 1,2}\\
    }
\affiliations {
    \textsuperscript{\rm 1}Anhui Province Key Laboratory of Big Data Analysis and Application,\\School of Computer Science and Technology, University of Science and Technology of China\\
    \textsuperscript{\rm 2}State Key Laboratory of Cognitive Intelligence\\
    \{yflyl613, wulk, yrunl, zaixi\}@mail.ustc.edu.cn, qiliuql@ustc.edu.cn, meet.leiyu@gmail.com
}

\begin{document}

\maketitle

\begin{abstract}
Federated recommendation (FedRec) can train personalized recommenders without collecting user data, but the decentralized nature makes it susceptible to poisoning attacks.
Most previous studies focus on the targeted attack to promote certain items, while the untargeted attack that aims to degrade the overall performance of the FedRec system remains less explored.
In fact, untargeted attacks can disrupt the user experience and bring severe financial loss to the service provider.
However, existing untargeted attack methods are either inapplicable or ineffective against FedRec systems.
In this paper, we delve into the untargeted attack and its defense for FedRec systems.
(i) We propose \textit{ClusterAttack}, a novel untargeted attack method.
It uploads poisonous gradients that converge the item embeddings into several dense clusters, which make the recommender generate similar scores for these items in the same cluster and perturb the ranking order.
(ii) We propose a \textbf{uni}f\textbf{o}rmity-based defe\textbf{n}se mechanism (\textit{UNION}) to protect FedRec systems from such attacks.
We design a contrastive learning task that regularizes the item embeddings toward a uniform distribution.
Then the server filters out these malicious gradients by estimating the uniformity of updated item embeddings.
Experiments on two public datasets show that \textit{ClusterAttack} can effectively degrade the performance of FedRec systems while circumventing many defense methods, and \textit{UNION} can improve the resistance of the system against various untargeted attacks, including our \textit{ClusterAttack}.
\end{abstract}

\section{Introduction}
Over recent years, personalized recommender systems have been widely used to alleviate the information overload problem for users~\cite{session2019,zhi2020visual,jin2020}.
Most conventional recommenders are trained on centralized user data, which has the risk of data leakage and raises privacy concerns~\cite{wu2018cover,fr2020}.
Moreover, some privacy regulations, such as GDPR\footnote{\url{https://gdpr-info.eu}} and CCPA\footnote{\url{https://oag.ca.gov/privacy/ccpa}}, make it more difficult to collect user data for centralized model training.
Federated learning (FL) is a decentralized training paradigm that enables multiple clients to collaboratively learn a global model without sharing their local data~\cite{fl2017}.
Several studies have applied FL to train privacy-preserving federated recommendation (FedRec) systems~\cite{fedrec2021,fedrec++2021}.

Unfortunately, FL is known to be vulnerable to poisoning attacks~\cite{adversarial2019, backdoorFL2020,provable2021}.
Its decentralized training procedure allows the attacker to arbitrarily modify the local training data or the uploaded gradient to achieve certain malicious goals.
Based on the goal of the attacker, poisoning attacks can be classified into targeted and untargeted attacks.
In the FedRec scenario, previous studies mainly focus on targeted attacks that try to promote certain target items~\cite{pipattack2022,fedrecattack2022}.
The untargeted attack that aims to degrade the overall performance of the FedRec system is still less explored~\cite{fedattack2022}.
In fact, without an effective defense mechanism, the untargeted attack can continuously disrupt the user experience of the system, which will lead to severe losses of customers and revenue for the service provider~\cite{cfpoi2016}.
Thus, it is necessary to explore the untargeted attack and its defense for FedRec systems.

The untargeted poisoning attack against FedRec systems faces several critical challenges.
First, the attack method must be effective even with a small fraction of malicious clients.
Considering that a recommender system usually has millions of users, it is impractical for the attacker to control a large number of clients.
Second, the attacker can only access a small set of data stored on the malicious clients as the clients never share their local training data in the FL framework.
Since existing poisoning attack methods against centralized recommendation model learning usually require a strong knowledge of the full training data~\cite{graphpoi2018,triple2021}, they are infeasible in the FedRec scenario.
Third, the untargeted attack aims to degrade the overall performance of FedRec systems on arbitrary inputs.
It is more challenging than the targeted attack that only manipulates the model output on certain target items.
Fourth, many recommenders are trained on implicit user feedback with heavy noise, which makes them robust to malicious perturbation to a certain degree~\cite{noisy2020,group2021}.

To address these challenges, in this paper, we first propose a novel untargeted model poisoning attack method named \textit{ClusterAttack}, which can effectively degrade the overall performance of FedRec systems with a small fraction of malicious clients.
Its main idea is to upload poisonous gradients that converge the item embeddings of the recommendation model into several dense clusters, which can let the recommender generate similar scores for these close items in the same cluster and perturb the ranking order.
Specifically, we split the item embeddings into several clusters with an adaptive clustering mechanism and compute the malicious gradient that reduces the within-cluster variance.
To make our attack more difficult to be detected, we clip the malicious gradient with a norm bound estimated from the normal gradients before uploading it to the server.
As most existing defense methods cannot effectively defend \textit{ClusterAttack}, we further propose a uniformity-based defense mechanism (\textit{UNION}) to protect FedRec systems from such attacks.
We require all benign clients to train the local recommendation model with an additional contrastive learning task that regularizes the item embeddings toward a uniform distribution in the space.
Then the server identifies these malicious gradients by estimating the uniformity of updated item embeddings.
In addition, our \textit{UNION} mechanism can be combined with many existing Byzantine-robust FL methods to provide more comprehensive protection for FedRec systems.
Extensive experiments on two public datasets show that our \textit{ClusterAttack} can effectively degrade the performance of FedRec systems without being detected, and our \textit{UNION} mechanism can improve the resistance of the system against many untargeted attacks, including our \textit{ClusterAttack}\footnote{Our code is available at \url{https://github.com/yflyl613/FedRec}.}.

The main contributions of our work are listed as follows:
\begin{itemize}[leftmargin=*]
    \item We propose \textit{ClusterAttack}, a novel untargeted model poisoning attack method, which reveals the security risk of FedRec systems even with existing defense methods.
    \item We propose \textit{UNION}, a defense mechanism that improves the resistance of FedRec systems against various untargeted poisoning attacks.
          To our best knowledge, it is the first defense mechanism specialized for FedRec systems.
    \item Extensive experiments on two public datasets validate the effectiveness of our \textit{ClusterAttack} and \textit{UNION} methods.
\end{itemize}

\section{Related Work}
\subsection{Attack \& Defense for Recommender Systems}
Poisoning attacks against recommender systems and their defense have been widely studied in the past decades~\cite{fun2004,svm2016,quick2019}.
However, these researches mainly focus on the centralized training of recommendation models.
They require the attacker or the server to have strong knowledge of the full training data to perform effective attacks or defenses, such as all user profiles~\cite{bandwagon2006} or the entire rating matrix~\cite{cfpoi2016}.
These methods are infeasible under the FL setting since the server cannot access the data of the clients.
Recently, some targeted poisoning attack methods have been proposed to boost certain target items in the FedRec scenario~\cite{pipattack2022,fedrecattack2022}, while the untargeted attack and its defense are still less explored.
\citet{fedattack2022} is a recent untargeted data poisoning attack against FedRec systems.
It subverts the local model training by choosing items closest to the user embedding as negative samples and the farthest ones as positive samples.
However, it only manipulates the input training data while keeping the local training algorithm unmodified, which limits its attack effect.
Our \textit{ClusterAttack} utilizes the vulnerability of FL and performs a more powerful model poisoning attack, which can effectively degrade the performance of the FedRec system with a small fraction of malicious clients.

\subsection{Attack \& Defense for Federated Learning}
In the general FL domain, several untargeted poisoning attack methods have been proposed and can be directly applied to degrade the performance of FedRec systems~\cite{fang2020,labelflip2020}.
For example, \citet{lie2019} propose to add a small amount of noise to each dimension of the average normal gradient, where the intensity of noise is estimated by the ratio of malicious clients.
\citet{fang2020} propose to perturb the uploaded gradient by adding noise in opposite directions inferred from normal updates.
However, these methods usually require a large fraction of malicious clients (e.g., $20\%$) to achieve a significant performance degradation, which is unrealistic for a FedRec system with millions of users.
To protect the FL system from potential poisoning attacks, researchers have also proposed several Byzantine-robust FL methods in the past few years~\cite{trimmedmean2018,fldetector2022}.
Although these defense methods can guarantee the convergence of the global model, we found that most of them perform poorly against carefully-designed poisoning attacks in the FedRec scenario.
Due to the diversity of user interests, the training data on each client is highly non-IID.
Some gradients uploaded by benign clients may also deviate from others, which makes it more difficult for the server to distinguish these malicious gradients.
Our \textit{UNION} mechanism can be combined with existing Byzantine-robust FL methods and improves their performance against many untargeted poisoning attacks, especially our \textit{ClusterAttack}.

\section{Preliminaries}
In this section, we introduce the settings of federated recommendation systems and the threat model used in this work.

\subsection{Federated Recommendation Systems}
Let $\mathcal{I}$ and $\mathcal{U}$ denote the sets of $M$ items and $N$ users/clients in a recommender system, respectively.
These clients try to train a global model collaboratively without sharing their private data.
We assume that the parameters of the recommendation model $\Theta$ consist of three parts: an item model $\Theta_{\operatorname{item}}$ that converts the item ID into the item embedding, a user model $\Theta_{\operatorname{user}}$ that infers the user interest embedding from the user profile (e.g., the user ID or historical interacted items), and a predictor model ${\Theta_{\operatorname{pred}}}$ that predicts a ranking score given an item embedding and a user embedding.
In each training round, the server first distributes the current global model parameters $\left[\Theta_{\operatorname{item}};\Theta_{\operatorname{pred}}\right]$ to $n$ randomly selected clients.
Then each selected client computes the update gradient $\mathbf{g}=\left[\mathbf{g}_{\operatorname{item}};\mathbf{g}_{\operatorname{user}};\mathbf{g}_{\operatorname{pred}}\right]$ with their local data.
Following previous works~\cite{ijcai2022,yu2022collaborative}, we assume they use BPR~\cite{bpr2009} with $L_2$ regularization to train the local model, i.e., the gradient $\mathbf{g}$ is generated by optimizing the following loss function:
\begin{equation}
    \mathcal{L}_{\operatorname{rec}}=-\operatorname{log}\left(\sigma\left(\hat{y}_{\operatorname{p}}-\hat{y}_{\operatorname{n}}\right)\right)+\lambda\left\lVert\Theta\right\rVert_2^2,
\end{equation}
where $\sigma$ is the sigmoid function.
$\hat{y}_{\operatorname{p}}$ and $\hat{y}_{\operatorname{n}}$ are the predicted ranking scores of the positive and negative items.
Next, the client uploads $\left[\mathbf{g}_{\operatorname{item}};\mathbf{g}_{\operatorname{pred}}\right]$ to the server and updates the local user model with $\mathbf{g}_{\operatorname{user}}$, which is not uploaded due to its privacy sensitivity~\cite{fairrec2021,wu2021hier}.
Finally, the server aggregates all the received gradients with certain aggregation rules and updates the global model.
Such training round proceeds iteratively until convergence.

\subsection{Threat Model}
\subsubsection{Attack Goal.}
The attacker aims to degrade the overall performance of the FedRec system on arbitrary inputs.
\subsubsection{Attack Capability and Knowledge.}
The attacker controls a set of malicious clients $\mathcal{U}_{\operatorname{mal}}$ which accounts for $m\%$ of $\mathcal{U}$.
As there are usually millions of users in a recommender system, we assume that $m$ should be small (e.g., $m=1$).
Following previous works~\cite{fedattack2022, pipattack2022}, we assume that the attacker has access to the training code, local model, and user data on the devices of malicious clients while cannot access the data or gradients of other benign clients.
The attacker can arbitrarily modify the gradients uploaded by the malicious clients.
We also assume the attacker does not know the aggregation rule used by the server.

\begin{figure}[t]
    \centering
    \includegraphics[width=\linewidth]{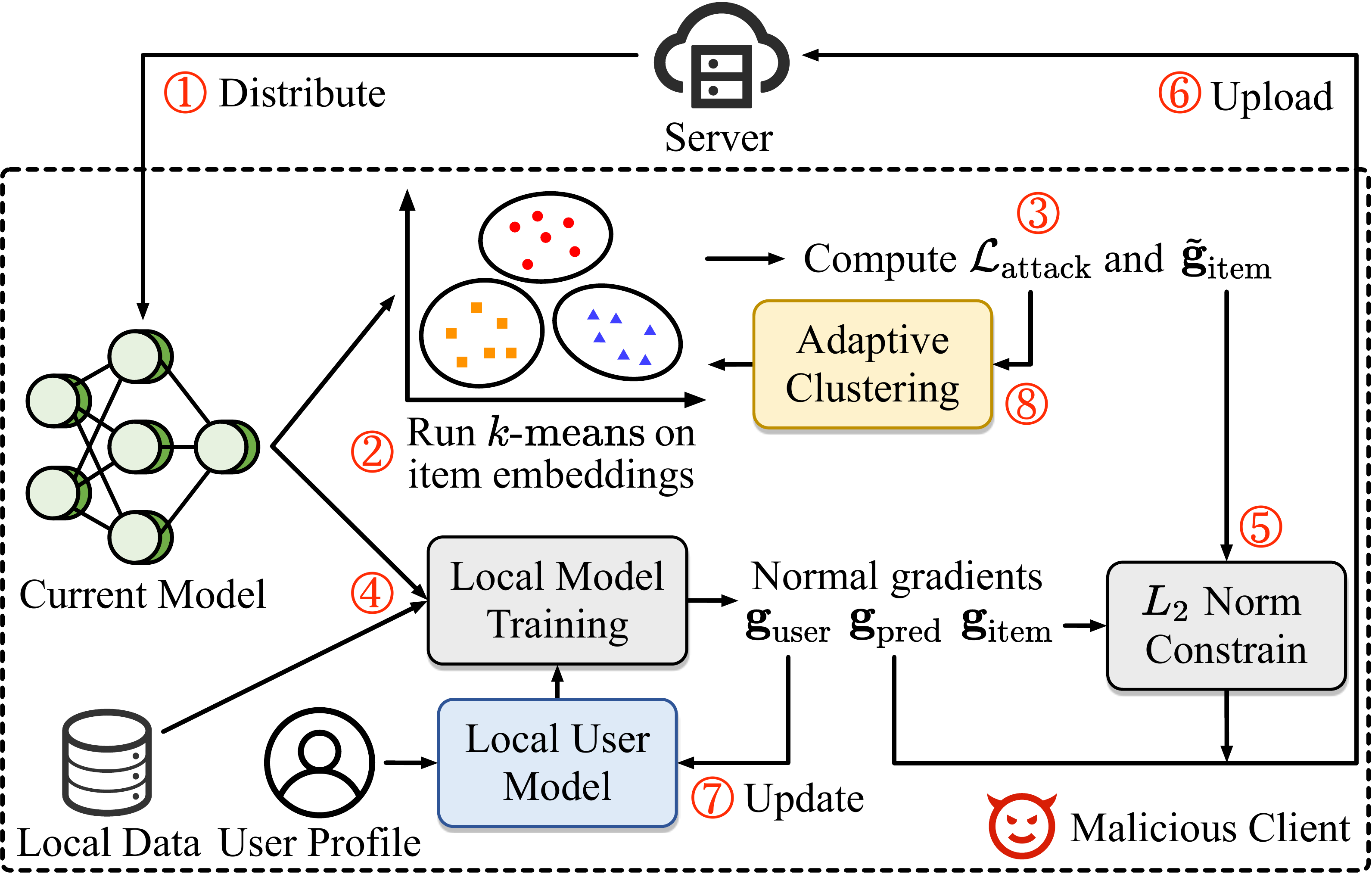}
    \caption{The procedure of our \textit{ClusterAttack}.}
    \label{clusterattack}
\end{figure}

\SetAlFnt{\small}
\begin{algorithm}[t]
    \SetAlgoNoLine
    \SetCommentSty{mycommfont}
    \caption{Adaptive Clustering}\label{alg:adaptive}
    \SetKwInOut{Input}{Input}
    \SetKwInput{Init}{Init}
    \SetKw{kwAnd}{and}
    \Input{Number of clusters $K$, range of number of clusters $\left[K_{\operatorname{min}}, K_{\operatorname{max}}\right]$, threshold $R$, and decay rate $\beta$.
    }
    \Init{Set $\tilde{\mathcal{L}}_{\operatorname{attack}}^{\sss (0)},n_{\operatorname{inc}},n_{\operatorname{dec}}$ and $t$ as 0.}
    \tcp{Repeat after each round of attack}
    $t \leftarrow t+1$\;
    Calculate $\mathcal{L}_{\operatorname{attack}}^{\sss (t)}$ with Equation~\eqref{loss_item}\;
    $\tilde{\mathcal{L}}_{\operatorname{attack}}^{\sss (t)}\leftarrow\beta\cdot\tilde{\mathcal{L}}_{\operatorname{attack}}^{\sss(t-1)}+(1-\beta)\cdot\mathcal{L}_{\operatorname{attack}}^{\sss (t)}$\;
    $\hat{\mathcal{L}}_{\operatorname{attack}}^{\sss (t)}\leftarrow\tilde{\mathcal{L}}_{\operatorname{attack}}^{\sss(t)}/\left(1-\beta^{t}\right)$\;
    \lIf{ $\hat{\mathcal{L}}_{\operatorname{attack}}^{\sss(t)}>\hat{\mathcal{L}}_{\operatorname{attack}}^{\sss(t-1)}\ $}{ $n_{\operatorname{inc}}\leftarrow n_{\operatorname{inc}}+1$}
    \lElse{ $n_{dec}\leftarrow n_{\operatorname{dec}}+1$}
    \If{ $n_{\operatorname{inc}}-n_{\operatorname{dec}}\geqslant R\ $}{$K\leftarrow\min\left(\left\lfloor K+\sqrt{K_{\operatorname{max}}-K}\right\rfloor, K_{\operatorname{max}}\right)$\;
    Reset $n_{\operatorname{inc}},n_{\operatorname{dec}}$ and $t$ as 0\;
    }
    \If{ $n_{\operatorname{dec}}-n_{\operatorname{inc}}\geqslant R\ $}{$K\leftarrow\max\left(\left\lfloor K-\sqrt{K-K_{\operatorname{min}}}\right\rfloor, K_{\operatorname{min}}\right)$\;
    Reset $n_{\operatorname{inc}},n_{\operatorname{dec}}$ and $t$ as 0\;}
\end{algorithm}

\section{Methodology}
In this section, we introduce the details of our untargeted model poisoning attack method \textit{ClusterAttack} and our defense mechanism \textit{UNION} for FedRec systems.

\subsection{\textit{ClusterAttack}}
To degrade the overall performance of FedRec systems, our \textit{ClusterAttack} aims to poison the item embeddings, which are widely used in most recommendation models~\cite{sasrec2018,ngcf2019}.
Since a recommendation model generally predicts the ranking score based on a user embedding and an item embedding, our main idea is to converge these item embeddings into several dense clusters.
Thus, the recommender tends to generate similar scores for these close items in the same cluster and mess up the ranking order.
Figure~\ref{clusterattack} illustrates the procedure of \textit{ClusterAttack}.

When selected for model training, the malicious client receives the latest global model from the server, which contains the item embeddings $\left\{\boldsymbol{v}_i\right\}_{i=1}^{\sss M}$ (Step 1).
We first apply $k$-means~\cite{kmeans1982} to split them into $K$ clusters $\left\{C_i\right\}_{i=1}^{\sss K}$ with centroids $\left\{\boldsymbol{c}_i\right\}_{i=1}^{\sss K}$ (Step 2).
Then we compute the following loss function to measure the within-cluster variance:
\begin{equation}
    \mathcal{L}_{\operatorname{attack}}=\sum_{i=1}^{K}\sum_{\boldsymbol{v}_j\in C_i}\left\lVert\boldsymbol{v}_j-\boldsymbol{c}_i\right\rVert_2^2.
    \label{loss_item}
\end{equation}
The malicious gradient of each item embedding is computed to minimize the above attack loss, i.e., $\boldsymbol{\tilde{g}}_{\boldsymbol{v}_i}=\partial{\mathcal{L}_{\operatorname{attack}}}/{\partial{\boldsymbol{v}_i}}$ (Step 3).
To make our attack stealthier, we further clip $\boldsymbol{\tilde{g}}_{\boldsymbol{v}_i}$ with an estimated norm of normal item embedding gradients.
Specifically, for each malicious client $u^{\sss (j)}\in \mathcal{U}_{\operatorname{mal}}$, we compute the normal gradient with the original loss function $\mathcal{L}_{\operatorname{rec}}$ and his local training data (Step 4).
Then we calculate the mean $\mu$ and standard deviation $\sigma$ of the $L_2$ norms of all normal item embedding gradients.
Assuming these norms follow a Gaussian distribution, we generate a reasonable norm bound $b_i^{\sss (j)}=\mu+\lambda_i^{\sss (j)}\sigma$ for each item embedding $\boldsymbol{v}_i$ on the malicious client $u^{\sss (j)}$, where $\lambda_i^{\sss (j)}$ is a number randomly sampled from $\left[0, 3\right]$.
Therefore, the clipped malicious item embedding gradients are formulated as follows:
\begin{equation}
    \hat{\mathbf{g}}_{\boldsymbol{v}_i}^{\sss (j)}=\frac{\tilde{\mathbf{g}}_{\boldsymbol{v}_i}}{\max\left(1, \left\lVert\tilde{\mathbf{g}}_{\boldsymbol{v}_i}\right\rVert_2/b_i^{\sss (j)}\right)}.
\end{equation}
The malicious gradient of the item model is set as $\hat{\mathbf{g}}_{\operatorname{item}}^{\sss (j)}=\left[\hat{\mathbf{g}}_{\boldsymbol{v}_1}^{\sss (j)};\hat{\mathbf{g}}_{\boldsymbol{v}_2}^{\sss (j)};\cdots;\hat{\mathbf{g}}_{\boldsymbol{v}_M}^{\sss (j)}\right]$.
Finally, the malicious client uploads $\hat{\mathbf{g}}^{\sss (j)}=\left[\hat{\mathbf{g}}_{\operatorname{item}}^{\sss (j)};\mathbf{g}_{\operatorname{pred}}^{\sss (j)}\right]$ to the server and updates its local user model with the normal gradient $\mathbf{g}_{\operatorname{user}}^{\sss (j)}$ (Step 6 \& 7).

Considering that the number of clusters $K$ will greatly impact the attack effect, we further design an adaptive clustering mechanism to automatically adjust the value of $K$ after each round of attack (Step 8), which is shown in \textbf{Algorithm~\ref{alg:adaptive}}.
Since the only feedback of the attack effect for the attacker is $\mathcal{L}_{\operatorname{attack}}$ in Equation~\eqref{loss_item}, we track it during the attack process and compute its bias-corrected exponential moving average $\hat{\mathcal{L}}_{attack}$.
We use two counters $n_{inc}$ and $n_{dec}$ to record the number of rounds in which the smoothed attack loss increases and decreases, respectively.
If $\hat{\mathcal{L}}_{attack}$ increased in most of the past few rounds, we assume that the current value of $K$ is too small, and the attack loss cannot converge well.
Thus, we increase the value of $K$ to make the attack easier.
Contrarily, if $\hat{\mathcal{L}}_{attack}$ keeps descending, we further decrease the value of $K$ to perform a stronger attack.

\SetAlFnt{\small}
\begin{algorithm}[!t]
    \SetAlgoLined
    \SetAlgoNoLine
    \caption{Defense Procedure on the Server Side}\label{server_defense}
    \SetKwInOut{Input}{Input}
    \Input{A set of received model gradients $\left\{\mathbf{g}^{\sss (i)}\right\}_{i=1}^n=\left\{\left[\mathbf{g}^{\sss (i)}_{\operatorname{item}};\mathbf{g}^{\sss (i)}_{\operatorname{pred}}\right]\right\}_{i=1}^n$, current global model $\boldsymbol{\Theta}=\left[\boldsymbol{\Theta}_{\operatorname{item}};\boldsymbol{\Theta}_{\operatorname{pred}}\right]$, learning rate $\eta$, the item set $\mathcal{I}$, and number of random sampling $T$.}
    \KwOut{A set of filtered model gradients $\mathcal{G}_{\operatorname{filter}}$.}
    \For{$i\leftarrow 1$ \KwTo $n$}{
    Update the item model $\boldsymbol{\Theta}_{\operatorname{item}}^{\sss (i)\prime}\leftarrow\boldsymbol{\Theta}_{\operatorname{item}}-\eta\cdot\mathbf{g}_{\operatorname{item}}^{\sss (i)}$\;
    Randomly select $T$ items $\left\{v_i\right\}_{i=1}^{\sss T}$ from $\mathcal{I}$ to estimate the uniformity of updated item embeddings $d_i\leftarrow\frac{2}{T(T-1)}\sum_{j=1}^{T}\sum_{k=j+1}^{T}\left\lVert f(v_j)-f(v_k) \right\rVert_2^2$\;
    }
    \eIf{$\operatorname{GapStatistics}(\left\{d_i\right\}_{i=1}^n)$}{
    Running $k$-means to split $\left\{d_i\right\}_{i=1}^n$ into two clusters\;
    $\mathcal{G}_{\operatorname{filter}}\leftarrow$ all the gradients in the larger cluster\;
    }{
    $\mathcal{G}_{\operatorname{filter}}\leftarrow\left\{\mathbf{g}^{\sss(i)}\right\}_{i=1}^{n}$\;
    }
    \KwRet{$\mathcal{G}_{\operatorname{filter}}$}
\end{algorithm}

\subsection{\textit{UNION} Mechanism}
Our \textit{ClusterAttack} reveals that maintaining the distribution of item embeddings is essential for protecting FedRec systems.
Thus, we further propose a uniformity-based defense mechanism (\textit{UNION}) that regularizes the item embeddings toward a uniform distribution in the space with a contrastive learning task.
Then the server filters out these malicious gradients that lead to abnormally distributed item embeddings.

\subsubsection{Client Side.}
We require all benign clients to train the local recommendation model with an additional contrastive learning (CL) task.
Specifically, denote the item set interacted by a client as $\mathcal{I}_u=\left\{v_i\right\}_{i=1}^{L}$.
For each $v_i\in\mathcal{I}_u$, we randomly select another item $v_i^\text{\scalebox{1}{+}}$ from $\mathcal{I}_u$ as the positive sample, and $P$ items $\left\{v_i^\text{\scalebox{1.5}{-}}\right\}_{i=1}^{\sss P}$ from $\mathcal{I}\backslash\mathcal{I}_u$ as the negative samples.
We adopt InfoNCE~\cite{infonce2018} as the contrastive loss function.
It is formulated as follows:
\begin{equation}
    \mathcal{L}_{\operatorname{cl}}=-\sum_{i=1}^{L}\log\frac{e^{f\left(v_i\right)^{\mathsf{T}}f\left(v_i^\text{\scalebox{1.1}{+}}\right)}}{e^{f\left(v_i\right)^{\mathsf{T}}f\left(v_i^\text{\scalebox{1.1}{+}}\right)}+\sum_{j=1}^P{e^{f\left(v_i\right)^{\mathsf{T}}f\left(v_j^\text{\scalebox{1.8}{-}}\right)}}},
\end{equation}
where $f$ denotes the item model.
The overall loss function on the client side is $\mathcal{L}=\mathcal{L}_{\operatorname{rec}}+\alpha\cdot\mathcal{L}_{\operatorname{cl}}$.
As proved by~\citet{align&uniform2020}, the contrastive loss $\mathcal{L}_{\operatorname{cl}}$ asymptotically optimizes the uniformity of the distribution induced from the learned embeddings, which is measured as follows:
\begin{equation}
    \mathcal{L}_{\operatorname{uniform}}(f;t) =\mathbb{E}_{x,y\overset{\text{i.i.d}}{\sim}p_{\operatorname{data}}}e^{-t\left\lVert f(x)-f(y)\right\rVert_2^2},\ \ t>0.\label{eq:uniform}
\end{equation}
Therefore, the additional CL task can regularize the item embeddings toward a uniform distribution in the space while training with the recommendation task.
Since such an optimization objective is opposite to the goal of \textit{ClusterAttack}, the CL task can mitigate its attack effect and also makes it easier for the server to distinguish these malicious gradients.

\begin{table*}[!t]\footnotesize
    \centering
    \begin{tabular}{cccccc}
        \toprule[1pt]
        \multirow{2}{*}[-2.5pt]{Model} & \multirow{2}{*}[-2.5pt]{\begin{tabular}[c]{@{}c@{}}Attack\\Method\end{tabular}} & \multicolumn{2}{c}{ML-1M} & \multicolumn{2}{c}{Gowalla}                                                           \\ \cmidrule(lr){3-4}\cmidrule(lr){5-6}
                                &                                                                          & HR@5                              & NDCG@5                      & HR@5                       & NDCG@5                     \\ \midrule
        \multirow{7}{*}{MF}     & No Attack                                                                & 0.03549 (-)                       & 0.02226(-)                  & 0.02523 (-)                & 0.01697 (-)                \\
                                & LabelFlip                                                                & 0.03561 (-0.34\%)                 & 0.02238 (-0.54\%)           & 0.02541 (-0.71\%)          & 0.01711 (-0.82\%)          \\
                                & FedAttack                                                                & 0.03358 (5.38\%)                  & 0.02118 (4.85\%)            & 0.02371 (6.02\%)           & 0.01585 (6.60\%)           \\
                                & Gaussian                                                                 & 0.03555 (-0.17\%)                 & 0.02224 (0.09\%)            & 0.02528 (-0.20\%)          & 0.01701 (-0.24\%)          \\
                                & LIE                                                                      & 0.03259 (8.17\%)                  & 0.02062 (7.37\%)            & 0.02316 (8.20\%)           & 0.01571 (7.42\%)           \\
                                & Fang                                                                     & 0.03038 (14.40\%)                 & 0.01897 (14.78\%)           & 0.02131 (15.54\%)          & 0.01448 (14.67\%)          \\
                                & ClusterAttack                                                            & \textbf{0.02451 (30.94\%)}        & \textbf{0.01545 (30.59\%)}  & \textbf{0.01664 (34.05\%)} & \textbf{0.01117 (34.18\%)} \\ \midrule
        \multirow{7}{*}{SASRec} & No Attack                                                                & 0.10810 (-)                       & 0.07053 (-)                 & 0.03251 (-)                        & 0.02217 (-)                        \\
                                & LabelFlip                                                                & 0.10857 (-0.43\%)                 & 0.07071 (-0.26\%)           & 0.03270 (-0.58\%)                  & 0.02222 (-0.23\%)                  \\
                                & FedAttack                                                                & 0.10013 (7.37\%)                  & 0.06572 (6.82\%)            & 0.03054 (6.06\%)                   & 0.02087 (5.86\%)                   \\
                                & Gaussian                                                                 & 0.10769 (0.38\%)                  & 0.07055 (-0.03\%)           & 0.03226 (0.77\%)                  & 0.02222 (-0.23\%)                  \\
                                & LIE                                                                      & 0.09677 (10.48\%)                 & 0.06281 (10.95\%)           & 0.03008 (7.47\%)                   & 0.02021 (8.84\%)                   \\
                                & Fang                                                                     & 0.08964 (17.08\%)                 & 0.05909 (16.22\%)           & 0.02797 (13.96\%)                  & 0.01883 (15.07\%)                  \\
                                & ClusterAttack                                                            & \textbf{0.06547 (39.44\%)}        & \textbf{0.04130 (41.44\%)}  & \textbf{0.02223 (31.62\%)}        & \textbf{0.01544 (30.36\%)}        \\ \bottomrule[1pt]
    \end{tabular}
    \caption{Model performance under different untargeted attack methods with no defense. The percentages in parentheses indicate the relative performance degradation compared with the no-attack scenario.}\label{attack_result}
\end{table*}

\SetAlFnt{\small}
\begin{algorithm}[!t]
    \SetAlgoLined
    \SetAlgoNoLine
    \caption{Gap Statistics}\label{gap_statistics}
    \SetKwInOut{Input}{Input}
    \Input{A set of estimated uniformity $\left\{d_{i}\right\}_{i=1}^{n}$, and number of sampling $B$.
    }
    \KwOut{Whether there is more than one cluster.}
    $\left\{\tilde{d}_i\right\}_{i=1}^n\leftarrow$ apply min-max normalization to $\left\{d_i\right\}_{i=1}^n$\;
    \For{$k\in\{1,2\}$ in parallel}{
    Apply $k$-means on $\left\{\tilde{d}_i\right\}_{i=1}^n$ to get clusters $\left\{D_i\right\}_{i=1}^k$ with centroids $\left\{\mu_i\right\}_{i=1}^k$\;
    $w_k \leftarrow\sum_{i=1}^k \sum_{\tilde{d}_j\in D_i}\left\lVert \tilde{d}_j-\mu_i \right\rVert_2^2$\;
    \For{$b\leftarrow 1$ \KwTo $B$}{
    Uniformly sample $n$ points $\left\{t_i\right\}_{i=1}^n$ from $[0,1]$\;
    Apply $k$-means on $\left\{t_i\right\}_{i=1}^n$ to get clusters $\left\{D^{\prime}_{i}\right\}_{i=1}^k$ with centoids $\left\{\mu^{\prime}_{i}\right\}_{i=1}^k$\;
    $w^*_{b} \leftarrow \sum_{i=1}^{k}\sum_{t_j\in D^{\prime}_i}\left\lVert t_j-\mu_i^{\prime} \right\rVert_2^2$\;
    }
    $\overline{w}=\frac{1}{B}\sum_{b=1}^{\sss B}\operatorname{log}\left(w^*_{b}\right)$\;
    $gap_k\leftarrow \overline{w}-\operatorname{log}\left(w_k\right)$\;
    $s_k\leftarrow\sqrt{\frac{1+B}{B^2}\sum_{b=1}^{\sss B}\left[\log(w^*_{b})-\overline{w}\right]^2}$\;
    }
    \KwRet{$gap_1<gap_2-s_2$}
\end{algorithm}

\subsubsection{Server Side.}
Since now all benign clients train the model with the CL task that optimizes the item embeddings toward a uniform distribution, we let the server estimate the uniformity of updated item embeddings for each received gradient.
Here we measure the uniformity in terms of the average $L_2$ distance between any two item embeddings, i.e.,
\begin{equation}
    \mathcal{L}_{\operatorname{uniform}}^{\prime}(f) =\mathbb{E}_{x,y\overset{\text{i.i.d}}{\sim}p_{\operatorname{data}}}\left\lVert f(x)-f(y)\right\rVert_2^2,
\end{equation}
which is a simplified version of Equation~\eqref{eq:uniform}.
The defense procedure on the server side is shown in \textbf{Algorithm~\ref{server_defense}}.
We further adopt the Gap Statistics algorithm~\cite{gapstatistics2001} to estimate the number of clusters in the set of estimated uniformity $\left\{d_i\right\}_{i=1}^n$, which is shown in \textbf{Algorithm~\ref{gap_statistics}}.
Generally, it compares the change of within-cluster dispersion with that expected under a uniform distribution to determine the number of clusters in a set of data.
If the algorithm estimates that there is more than one cluster, we believe there are some malicious gradients that lead to abnormally distributed item embeddings.
Hence, we further apply $k$-means to split $\left\{d_i\right\}_{i=1}^n$ into two clusters and filter out all the gradients belonging to the minor one.

It is noted that \textit{UNION} is a general mechanism that aims to preserve the distribution of item embeddings.
It can be easily combined with many existing Byzantine-robust FL methods~\cite{krum2017,normbound2020} to provide more comprehensive protection for FedRec systems.
These methods can learn a more accurate model on the set of filtered model gradients returned by our \textit{UNION} mechanism while maintaining their original convergence guarantee.

\section{Experiments}
In this section, we conduct several experiments to answer the following research questions (RQs):
\begin{itemize}[leftmargin=*]
    \item \textbf{RQ1}: How does our \textit{ClusterAttack} perform compared with existing untargeted attack methods?
    \item \textbf{RQ2}: Can our \textit{ClusterAttack} circumvent existing defense methods while preserving its attack performance?
    \item \textbf{RQ3}: How does our \textit{UNION} mechanism perform against existing untargeted attacks and our \textit{ClusterAttack}?
    \item \textbf{RQ4}: How does the ratio of malicious clients affect the performance of our methods?
    \item \textbf{RQ5}: Is the proposed adaptive clustering mechanism in our \textit{ClusterAttack} effective?
    \item \textbf{RQ6}: How difficult it is to defend our \textit{ClusterAttack}, and why does our \textit{UNION} mechanism work?
\end{itemize}

\subsection{Datasets and Experimental Settings}
We conduct experiments with two public datasets.
The first is ML-1M~\cite{movielens2016}, a movie recommendation dataset.
The second is Gowalla~\cite{gowalla2016}, a check-in dataset obtained from the Gowalla website.
We adopt the 10-core version used in~\cite{ngcf2019}, i.e., retaining users and items with at least ten interactions.
The statistics of the two datasets are shown in Table~\ref{stat}.
Following previous works~\cite{ncf2017,bert4rec2019}, we adopt the leave-one-out approach and hold out the latest interacted item of each user as the test data.
We use the item before the last one for validation and the rest for training.

In our experiments, we choose the widely used MF~\cite{bpr2009} and SASRec~\cite{sasrec2018} as the recommendation model.
The hidden dimension of models is 64.
We use FedAvg~\cite{fl2017} with Adam optimizer~\cite{adam2015} as the FL framework.
Each user is treated as a client in the FedRec system.
50 clients are randomly selected in each round for model training.
We randomly select $1\%$ of users from the entire user set $\mathcal{U}$ and take them as malicious clients.
The detailed experimental settings are listed in the Appendix.
Following~\cite{pipattack2022, fedattack2022}, we use the Hit Ratio (HR) and the Normalized Discounted Cumulative Gain (NDCG) over the top 5 ranked items to measure the performance of the recommendation model.
Note that the metrics are only calculated on benign clients using the all-ranking protocol, i.e., all items not interacted with by the user are used as candidates.
All the hyper-parameters are tuned on the validation set.
We repeat each experiment 5 times and report the average results.

\begin{table}[!t]
    \centering
    \resizebox{\linewidth}{!}{
        \begin{tabular}{cccccc}
            \toprule[1pt]
            Dataset & \#Users & \#Items & \#Actions & Avg. length & Density \\ \midrule
            ML-1M   & 6,040   & 3,706   & 1,000,209 & 165.6       & 4.47\%  \\
            Gowalla & 29,858  & 40,981  & 1,585,043 & 53.1        & 0.13\%  \\
            \bottomrule[1pt]
        \end{tabular}
    }
    \caption{Detailed statistics of the two datasets.}\label{stat}
\end{table}

\begin{figure*}[!t]
    \centering
    \includegraphics[width=\linewidth]{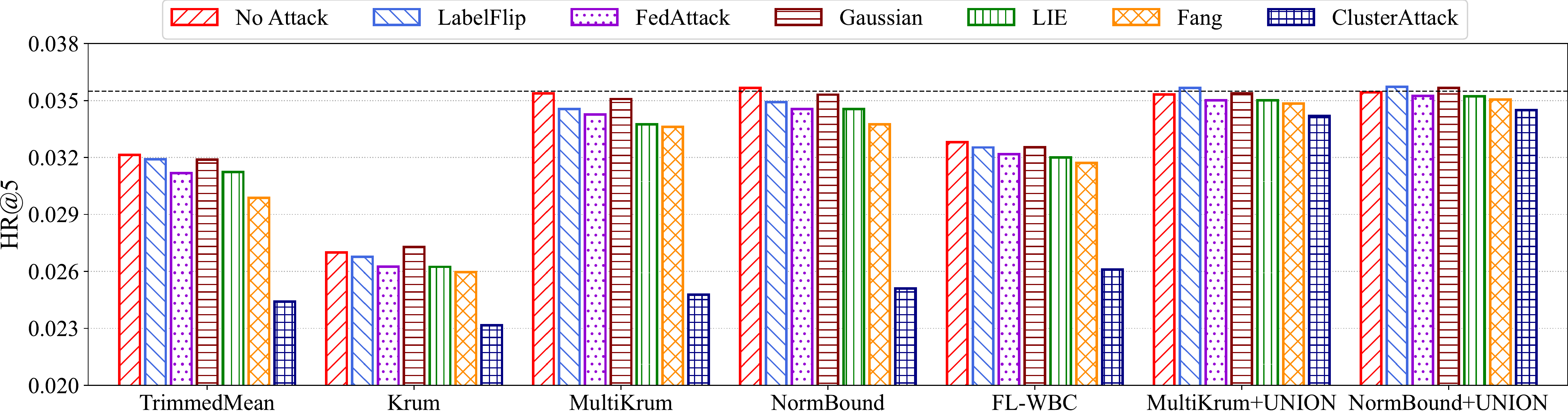}
    \caption{Model performance under different untargeted attack methods with different defense mechanisms. The black dashed line represents the model performance without any attack or defense.}
    \label{res:defense}
\end{figure*}

\subsection{Attack Performance Evaluation (RQ1)}
We compare the attack performance of \textit{ClusterAttack} with the following \textit{data poisoning attack} methods:
\begin{itemize}[leftmargin=*]
    \item \textit{LabelFlip}~\cite{labelflip2020}, which flips the label of the training sample on malicious clients;
    \item \textit{FedAttack}~\cite{fedattack2022}, which chooses items that are most similar to the user embedding as negative samples and the farthest ones as positive samples;
\end{itemize}
and the following \textit{model poisoning attack} methods:
\begin{itemize}[leftmargin=*]
    \item \textit{Gaussian}~\cite{fang2020}, which computes the mean and standard deviation of normal gradients and then uploads samples from the estimated Gaussian distribution;
    \item \textit{LIE}~\cite{lie2019}, which adds small amounts of noise to the average normal gradient;
    \item \textit{Fang}~\cite{fang2020}, which adds noise in opposite directions to the average normal gradient.
\end{itemize}

The experimental results are shown in Table~\ref{attack_result}.
From the results, we have several findings.
First, not all untargeted poisoning attacks are effective against FedRec systems when the number of malicious clients is limited.
Especially, \textit{LabelFlip} and \textit{Gaussian} even slightly raise the performance of the model.
This may be because such limited perturbations make the recommendation model more robust to the noise in user behaviors.
Second, well-designed model poisoning attacks (\textit{LIE}, \textit{Fang}, and \textit{ClusterAttack}) usually perform better than data poisoning attacks (\textit{LabelFlip} and \textit{FedAttack}).
This is because these model poisoning attacks directly modify the uploaded gradient, which is more flexible and effective than manipulating the training data.
Third, our \textit{ClusterAttack} consistently outperforms other baselines by a large margin.
The reason is that our method uploads the poisonous gradient that converges the item embeddings into several dense clusters.
It is more effective than these baseline model poisoning attacks that only add certain perturbation noise to the normal gradients.
Besides, our adaptive clustering mechanism can automatically adjust the number of clusters after each round of attack, which also leads to better attack performance.

\subsection{Attack Effectiveness under Defense (RQ2)}
To verify the effectiveness of \textit{ClusterAttack} under defense, we compare the attack performance of different attack methods against the following Byzantine-robust FL methods:
\begin{itemize}[leftmargin=*]
    \item \textit{TrimmedMean}~\cite{trimmedmean2018}, which computes the coordinate-wise trimmed mean of the received gradients.
    \item \textit{Krum}~\cite{krum2017}, which selects the gradient closest to its neighboring gradients for the model update.
    \item \textit{MultiKrum}~\cite{krum2017}, which selects multiple gradients via \textit{Krum} and calculates their average.
    \item \textit{NormBound}~\cite{normbound2020}, which clips the $L_2$ norm of received gradients with a threshold before aggregation.
    \item \textit{FL-WBC}~\cite{sun2021flwbc}, in which all benign clients upload partially masked gradients with Laplace noise to mitigate the attack effect on the global model.
\end{itemize}

We use MF and ML-1M in the following experiments (the results with SASRec and Gowalla show similar trends and are omitted due to the space limit).
The results are shown as the \emph{left five} groups in Figure~\ref{res:defense}.
We find that some Byzantine-robust FL methods such as \textit{TrimmedMean} and \textit{Krum} severely degrade the performance of the recommendation model even without any attack.
We attribute it to the highly non-IID training data on each client.
Some benign gradients may also deviate from others.
Therefore, they may be incorrectly filtered out by these defense methods, which will impair the performance of the global model.
The partially masked gradients used in \textit{FL-WBC} will cause certain information loss which also leads to performance degradation.
In contrast, \textit{MultiKrum} and \textit{NormBound} do not substantially hurt the model performance and can mitigate the impact of most existing attacks.
However, none of these existing defense methods can effectively defend \textit{ClusterAttack}.
The reason is that we only manipulate the item embedding gradients and further bound their $L_2$ norms by the one estimated from normal gradients, which makes it less likely to be detected as an outlier by these defense methods.
Besides, our further analysis finds that the non-IID training data on each client also covers our attack (See Section~\ref{sec:rq6}).

\subsection{Defense Performance Evaluation (RQ3)}
\begin{table}[!t]
    \centering
    \resizebox{\linewidth}{!}{
        \begin{tabular}{ccc}
            \toprule[1pt]
            Defense Method                   & Attack Method    & HR@5              \\ \midrule
            \multirow{2}{*}{MultiKrum+UNION} & ClusterAttack    & 0.03378 (4.82\%)  \\
                                             & ClusterAttack+CL & 0.03525 (0.68\%)  \\\midrule
            \multirow{2}{*}{NormBound+UNION} & ClusterAttack    & 0.03449 (2.82\%)  \\
                                             & ClusterAttack+CL & 0.03566 (-0.48\%) \\ \bottomrule[1pt]
        \end{tabular}
    }
    \caption{Attack performance of \textit{ClusterAttack+CL}.}\label{attackcl}
\end{table}
In this subsection, we evaluate the effectiveness of our \textit{UNION} mechanism.
Since \textit{MultiKrum} and \textit{NormBound} will not greatly hurt the model performance, we combine them with our \textit{UNION} mechanism respectively and test their defense performance against different attacks.
The results are shown as the \emph{right two} groups in Figure~\ref{res:defense}.
We first find that our \textit{UNION} mechanism can significantly improve the resistance of these defense methods against \textit{ClusterAttack}.
The reason is that the additional CL task optimizes the item embeddings toward a uniform distribution, which is opposite to the goal of \textit{ClusterAttack}.
Besides, the server also keeps filtering out the gradients leading to abnormally distributed item embeddings.
Thus, the goal of \textit{ClusterAttack} cannot be effectively achieved.
We also find that our \textit{UNION} mechanism enhances the performance of these defense methods against other baseline attacks.
This verifies that regularizing the distribution of item embeddings is beneficial to protecting FedRec systems from various untargeted attacks.

Since our \textit{UNION} mechanism modifies the training algorithm on the client side, which can be known by the attacker via malicious clients, we wonder whether the attacker can avoid detection by generating malicious gradients along with the CL task.
Thus, we further evaluate the defense performance of \textit{UNION} against a new attack method \textit{ClusterAttack+CL}, which generates the malicious gradient by optimizing $\mathcal{L}_{\operatorname{attack}}^{\prime}=\mathcal{L}_{\operatorname{attack}}+\alpha\cdot\mathcal{L}_{\operatorname{cl}}$.
As shown in Table~\ref{attackcl}, the extra CL task weakens the attack effect of \textit{ClusterAttack}.
This is because $\mathcal{L}_{\operatorname{attack}}$ tries to converge the item embeddings into clusters, while $\mathcal{L}_{\operatorname{cl}}$ regularizes these embeddings to be uniformly distributed.
Such opposite goals cannot be jointly optimized well and lead to poor attack effects.

\subsection{Influence of the Ratio of Malicious Clients (RQ4)}
\begin{table}[t]
    \resizebox{\linewidth}{!}{
        \begin{tabular}{cccccc}
            \toprule[1pt]
            $m\%$ & No Attack & FedAttack & LIE     & Fang    & ClusterAttack \\ \midrule
            $0.5\%$ & 0.03549   & 0.03491   & 0.03465 & 0.03426 & 0.03001       \\
            $1\%$   & 0.03549   & 0.03358   & 0.03259 & 0.03038 & 0.02451       \\
            \bottomrule[1pt]
        \end{tabular}
    }
    \caption{Attack performance of different untargeted attacks with different ratios of malicious clients.}\label{few-attack}
\end{table}

\begin{table}[t]
    \resizebox{\linewidth}{!}{
        \begin{tabular}{ccccc}
            \toprule[1pt]
            Defense Method  & FedAttack & LIE     & Fang    & ClusterAttack \\ \midrule
            No Defense      & 0.03195   & 0.03147 & 0.02793 & 0.01950       \\
            MultiKrum+UNION & 0.03438   & 0.03447 & 0.03454 & 0.03291       \\
            MormBound+UNION & 0.03490   & 0.03464 & 0.03464 & 0.03351       \\
            \bottomrule[1pt]
        \end{tabular}
    }
    \caption{Defense performance of \textit{UNION} against different untargeted attacks with $5\%$ malicious clients.}\label{more-attack}
\end{table}
In this subsection, we further conduct several experiments to explore how the ratio of malicious clients affects the performance of our methods.
First, we set the ratio of malicious clients $m\%$ as $0.5\%$ and $1\%$ respectively and compare the performance of our \textit{ClusterAttack} with various baseline untargeted attacks.
The HR@5 of the model are shown in Table~\ref{few-attack}. 
It shows that most existing attack methods are ineffective with very few malicious clients, while our \textit{ClusterAttack} can still degrade the model performance by $15.49\%$ even with $0.5\%$ malicious clients.
This verifies that our malicious gradients which aim to converge the item embeddings into dense clusters are highly effective in perturbing the ranking order of the FedRec system.
Next, to validate the robustness of our \textit{UNION} mechanism, we increase the ratio of malicious clients to $5\%$ and evaluate its performance against various attacks.
The HR@5 of the model shown in Table~\ref{more-attack} demonstrate that after combining with our \textit{UNION} mechanism, \textit{MultiKrum+UNION} and \textit{NormBound+UNION} can well protect the FedRec system against various untargeted attacks even with a large number of malicious clients.

\subsection{Impact of Adaptive Clustering (RQ5)}\label{sec:adaptive}
\begin{figure}[!t]
    \centering
    \includegraphics[width=\linewidth]{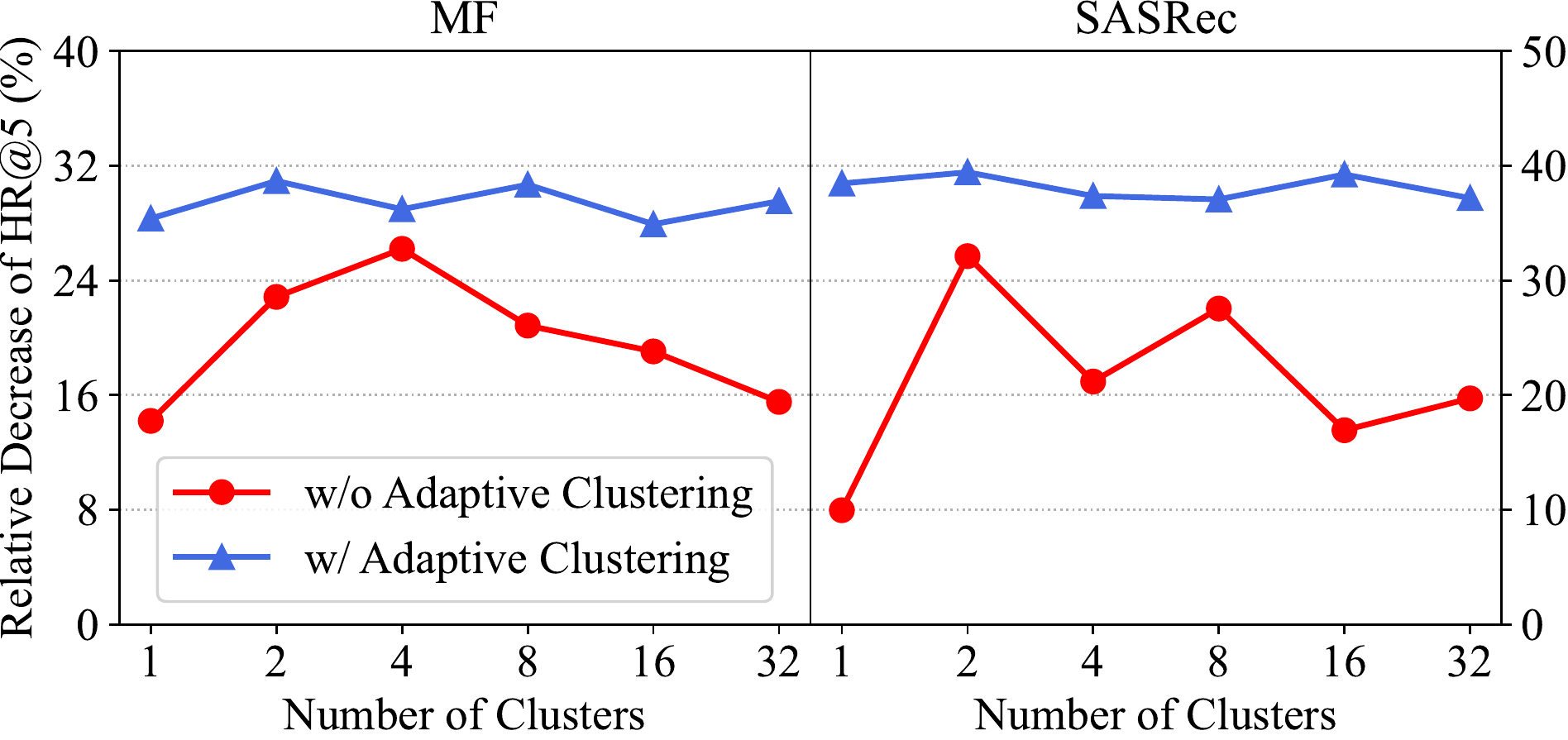}
    \caption{Impact of adaptive clustering.}
    \label{res:adaptive}
\end{figure}

In this subsection, we conduct experiments to verify the impact of the adaptive clustering mechanism in \textit{ClusterAttack}.
We set the initial number of clusters $K$ as $\{1,2,4,8,16\}$ respectively and compare the attack performance of \textit{ClusterAttack} and its variant with the adaptive clustering mechanism removed.
The results on the ML-1M dataset are shown in Figure~\ref{res:adaptive}.
We find that without the adaptive clustering mechanism, the attack effect does not always improve with the decreasing number of clusters as intuitively thought.
This is because when the value of $K$ is too small, the attack loss cannot effectively converge with limited malicious clients.
Results also show that the attack effect varies significantly with different numbers of clusters.
The attack performance is consistently better when the adaptive clustering mechanism is adopted to adjust the value of $K$ based on the attack effect after each round of attack, and it is less influenced by the selection of the initial number of clusters.

\subsection{Gradients and Uniformity Analysis (RQ6)}\label{sec:rq6}
\begin{figure*}[!t]
    \centering
    \includegraphics[width=\linewidth]{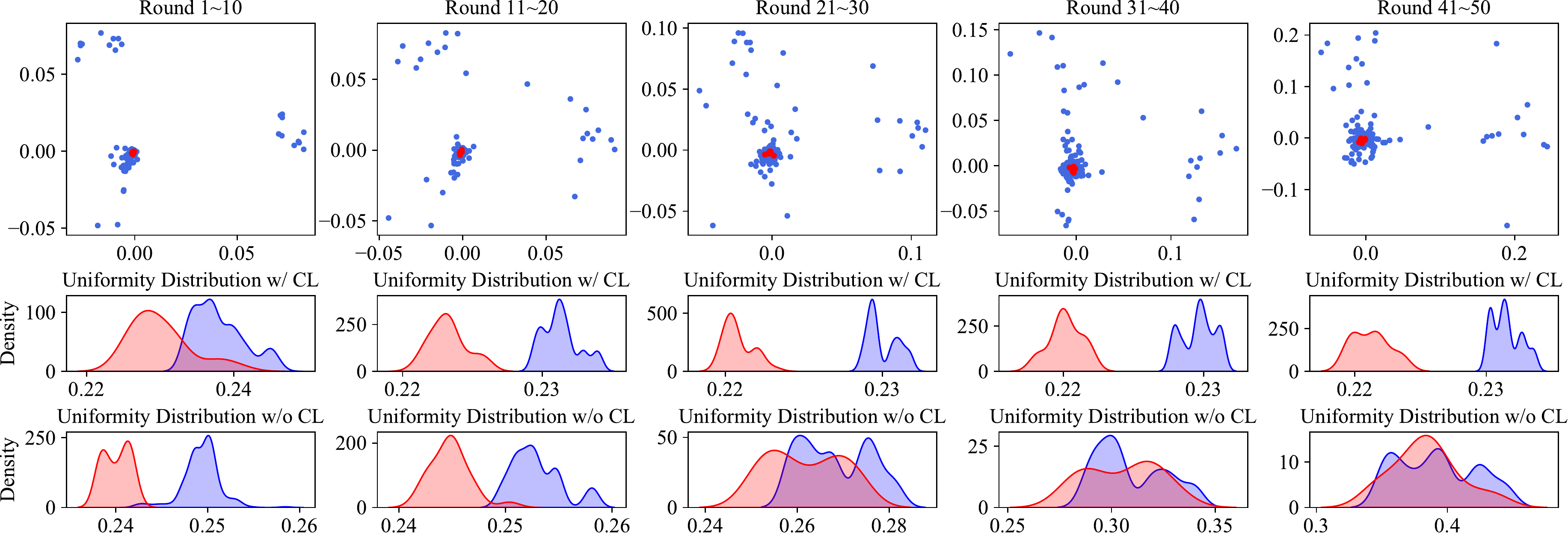}
    \caption{Visualization of the uploaded gradients and the uniformity distribution in different rounds of model training. The blue color and red color denote benign clients and malicious clients, respectively.}
    \label{visualize}
\end{figure*}
To better understand how difficult it is to defend \textit{ClusterAttack}, we save all the model gradients received by the server during the training procedure.
Every 10 rounds, we use PCA to reduce the dimension of these saved gradients and visualize them in the first row of Figure~\ref{visualize}.
We find that these malicious gradients have similar PCA components to most normal gradients, while some gradients uploaded by benign clients seem to be outliers.
This verifies that it is quite difficult to distinguish these malicious gradients due to the highly non-IID training data on each client.
We further save all the item embedding uniformities estimated by the server with our \textit{UNION} mechanism and its variant when the CL task on the client side is removed.
We use kernel density estimation (KDE)~\cite{kde1962} to estimate their probability density distributions, which are visualized in the last two rows of Figure~\ref{visualize}.
It shows that without the CL task, the item embeddings updated with normal gradients and malicious gradients tend to have similar uniformity.
In contrast, the estimated uniformities fall into two well-separated parts with the CL task.
Therefore, the server can better filter out these malicious gradients and train a more accurate model.

\section{Conclusion}
In this paper, we studied the untargeted poisoning attack and its defense for FedRec systems.
We first presented a novel untargeted attack named \textit{ClusterAttack}.
By uploading poisonous gradients that converge the item embeddings into several dense clusters, it can effectively degrade the overall performance of the FedRec system.
Meanwhile, we proposed a uniformity-based defense mechanism to protect the system from such attacks.
It requires all clients to train the recommendation model with an additional contrastive learning task, which enables the server to distinguish these malicious gradients based on the estimated uniformity of updated item embeddings.
Extensive experiments validated the effectiveness of our attack and defense methods.
Our work reveals the security risk of FedRec systems and provides a general defense mechanism that can be combined with existing Byzantine-robust FL methods to better protect the system from potential untargeted attacks in the real world.

\begin{appendices}
\section{Experimental Settings}
In our experiments, the hidden dimensions of both MF and SASRec models are set to 64.
We set the dropout ratio as 0.2 to mitigate overfitting.
The maximum sequence length of SASRec is set to 200 and 100 for ML-1M and Gowalla, respectively.
We use FedAvg~\cite{fl2017} with Adam optimizer~\cite{adam2015} as the FL framework.
The FL training procedure runs for at most 6,000 rounds to ensure the convergence of the global recommendation model.
The number of clients randomly selected per round $n$ is 50.
In our \textit{ClusterAttack}, we set the initial number of clusters as 2 and 8 for ML-1M and Gowalla respectively considering the size of their item sets.
The range of the number of clusters and the threshold $R$ is set to $[1, 50]$ and 100 for both datasets.
In our \textit{UNION} mechanism, we set the number of negative samples $P$ in the contrastive learning task as 15.
The coefficient $\alpha$ is set to 1.
The number of random sampling for uniformity estimation $T$ is 500.
In the Gap Statistics algorithm, the number of random sampling $B$ is 50.
The hyper-parameter settings specific to each model and dataset are listed in Table~\ref{hyper-parameter}.

\begin{table}[!t]
    \centering
    \resizebox{\linewidth}{!}{
        \begin{tabular}{ccccc}
            \toprule[1pt]
            \multicolumn{1}{c}{\multirow{2}{*}[-1.5pt]{Hyper-parameters}} & \multicolumn{2}{c}{ML-1M} & \multicolumn{2}{c}{Gowalla}                 \\ \cmidrule(lr){2-3}\cmidrule(lr){4-5}
            \multicolumn{1}{c}{}                                          & MF                        & SASRec                      & MF   & SASRec \\ \midrule
            Learning rate $\eta$                                          & 2e-3                      & 1e-3                        & 2e-3 & 1e-3   \\
            $L_2$ regularization coefficient $\lambda$                    & 1e-5                      & 1e-5                        & 1e-6 & 1e-5   \\
            $L_2$ norm bound for \textit{NormBound}                       & 0.1                       & 1.5                         & 0.1  & 2.5    \\ \bottomrule[1pt]
        \end{tabular}
    }
    \caption{Hyper-parameter settings.}\label{hyper-parameter}
\end{table}

\section{Experimental Environment}
We conduct experiments on a Linux server with CentOS 7.9.2009.
All experiments were run on an NVIDIA GeForce RTX 3090 GPU with CUDA 11.0.
The CPU is Intel(R) Xeon(R) Gold 6226R CPU @2.90GHz and the total memory is 376GB.
We use Python 3.9.12 and PyTorch 1.7.1.
\end{appendices}

\section*{Acknowledgements}

This research was partially supported by grants from the National Key Research and Development Program of China (No. 2021YFF0901003), and the National Natural Science Foundation of China (No. 61922073 and U20A20229).

\bibliography{aaai23}

\end{document}